\documentclass[pdflatex,sn-mathphys-num]{sn-jnl-mod}

\usepackage{graphicx}%
\usepackage{multirow}%
\usepackage{amsmath,amssymb,amsfonts}%
\usepackage{amsthm}%
\usepackage{mathrsfs}%
\usepackage[title]{appendix}%
\usepackage{xcolor}%
\usepackage{textcomp}%
\usepackage{manyfoot}%
\usepackage{booktabs}%
\usepackage{algorithm}%
\usepackage{algorithmicx}%
\usepackage{algpseudocode}%
\usepackage{listings}%

\begin{document}

\title[The Nonabelian Plasma is Chaotic]{The Nonabelian Plasma is Chaotic}
\subtitle{Egenstate Thermalization in SU(2) Gauge Theory}

\author*[1]{\fnm{Berndt} \sur{M\"uller}}\email{bmueller@duke.edu}
\equalcont{These authors contributed equally to this work.}

\author[2]{\fnm{Lukas} \sur{Ebner}}\email{Lukas.Ebner@physik.uni-regensburg.de}
\equalcont{These authors contributed equally to this work.}

\author[2]{\fnm{Andreas} \sur{Sch\"afer}}\email{Andreas.Schaefer@physik.uni-regensburg.de}
\equalcont{These authors contributed equally to this work.}

\author[2]{\fnm{Clemens} \sur{Seidl}}\email{Clemens.Seidl@physik.uni-regensburg.de}
\equalcont{These authors contributed equally to this work.}

\author[3]{\fnm{Xiaojun} \sur{Yao}}\email{xjyao@uw.edu}
\equalcont{These authors contributed equally to this work.}

\affil[1]{\orgdiv{Department of Physics}, \orgname{Duke University}, \city{Durham}, \postcode{27708}, \state{NC}, \country{USA}}

\affil*[2]{\orgdiv{Institut f\"ur Theoretische Physik}, \orgname{Universit\"at Regensburg}, \city{Regensburg}, \postcode{D-93040}, \country{Germany}}

\affil[3]{\orgdiv{
InQubator for Quantum Simulation, Department of Physics}, \orgname{University of Washington}, \city{Seattle}, \postcode{98105}, \state{WA}, \country{USA}}

\date{August 2024}

\abstract{Nonabelian gauge theories are chaotic in the classical limit. We discuss new evidence from SU(2) lattice gauge theory that they are also chaotic at the quantum level. We also describe possible future studies aimed at discovering the consequences of this insight. Based on a lecture presented by the first author at the {\it Particles and Plasmas} Symposium 2024.}


\pacs{IQuS@UW-21-086}

\maketitle

\section{Introduction}\label{sec:Intro}

The nonabelian lattice gauge theory at the classical level has long been known to be a chaotic dynamical system with ergodic properties \cite{Muller:1992iw,Biro:1993qc,Bolte:1999th}. The interesting and more challenging question is whether these gauge theories are also chaotic systems at the quantum level. A quantitative answer to this question would allow us to better understand how nonabelian gauge fields thermalize and what the relevant time scales for the various stages of thermalization are. After briefly reviewing the evidence for classical chaos of SU(2) gauge theory, we will discuss recent results showing that the quantum theory satisfies the Eigenstate Thermalization Hypothesis (ETH) and present results for the entanglement entropy of subsystems as function of the energy.

\section{Classical chaos for nonabelian gauge theories}
\label{sec:class}

The chaotic nature of a classical dynamical system is revealed by the exponential rate of divergence of neighboring trajectories in phase space. This rate is called the Lyapunov exponent. Conservative dynamical systems, such as classical nonabelian lattice gauge fields (at the classical level), with $2N$ degrees of freedom have up to $2(N-1)$ nonvanishing Lyapunov exponents that come in pairs of opposite signs. In the case of gauge theories, the gauge degrees of freedom are not dynamical and associated with vanishing Lyapunov exponents. The sum of all positive Lyapunov exponents, $S_{\rm KS}=\sum_{\lambda_k>0}\lambda_k$ is known as the Kolmogorov-Sina{\"i} entropy rate. It describes the growth rate of the coarse-grained entropy of an ensemble.

The spectrum of Lyapunov exponents for the (3+1)-dimensional pure SU(2) and SU(3) lattice gauge theory has been computed numerically starting around 1990. A systematic review of this topic can be found in \cite{Biro:1994bi}. The main results are as follows: 
\begin{figure}[h!]
\centering
\includegraphics[width=0.45\linewidth]{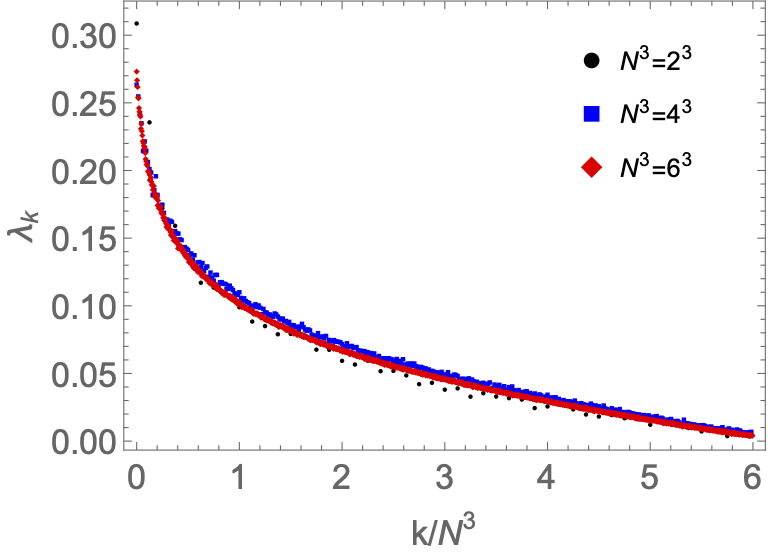}
\includegraphics[width=0.45\linewidth]{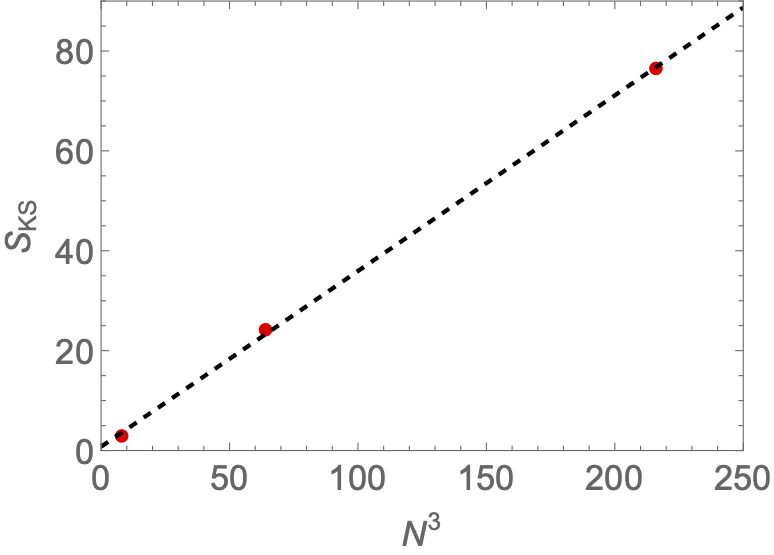}
\caption{Left panel: Spectrum of positive Lyapunov exponents $\lambda_k$ of SU(2) gauge theory, ordered by size, for cubic lattices of size $N^3$ with $N=2,4,6$ \cite{Bolte:1999th}. The abscissa shows the index, $1\leq k\leq 6N^3$, of the exponent $k$ scaled by $N^3$ . Right panel: KS-entropy rate $S_{\rm KS}$ versus lattice volume $N^3$. The linear relationship indicates that the lattice gauge theory is characterized by a linear growth rate of the coarse-grained entropy density.}
\label{fig:lyap}
\end{figure}
\begin{enumerate}
\item The maximal Lyapunov exponent scales linearly in the dimensional quantity $g^2E_Pa$, where $g$ is the coupling constant, $E_P$ is the average energy per plaquette, and $a$ is the lattice spacing. After expressing $E_P$ with the corresponding temperature $T$ at equilibrium, the maximal Lyapunov exponents for SU(2) and SU(3) can be combined into the formaula $\lambda_{\rm max}\approx 0.17g^2N_cT$, where $N_c$ denotes the number of colors. It turns out that this expression is equal to twice the thermal damping rate of a plasmon at rest \cite{Biro:1994sh,Braaten:1990it}
\item The lattice gauge field on a $N^3$ lattice has the maximally allowed number $2(N_c^2-1)N^3$ of positive Lyapunov exponents, which scale when plotted against the volume scaled index $k/N^3$ (see left panel of Fig.~\ref{fig:lyap}). This result implies that the KS-entropy rate $S_{\rm KS}$ is an extensive quantity (right panel of Fig.~\ref{fig:lyap}). In other words, the gauge field exhibits a linear growth rate for the coarse-grained entropy density \cite{Bolte:1999th}. It is remarkable that this behavior is already clearly seen for such small lattices.
\end{enumerate}
Maldacena, Shenker, and Stanford \cite{Maldacena:2015waa} (MSS) have argued that there exists an upper bound on Lyapunov exponents, $\lambda \leq 2\pi T$, where $T$ is the temperature reached after equilibration. Our numerical simulations for the SU(3) gauge theory \cite{Gong:1992yu} found that $\lambda_{\rm max}\approx 0.53g^2T$, which saturates the MSS bound when $\alpha_s =g^2/4\pi \approx 1$. The conclusion from these investigations is that the classical SU(3) gauge field is fully chaotic and ranks among the most strongly chaotic systems possible. Such systems are sometimes called ``fast scramblers'' \cite{Sekino:2008he}.

\section{Quantum chaos of nonabelian gauge theories}
\label{sec:quant}

As classical trajectories do not exist in quantum mechanics, one needs a different criterion to identify chaotic behavior. One such indicator are out-of-time correlators of noncommuting operators (OTOC) \cite{Larkin:1969xx,Garcia-Mata:2022voo}, whose exponential growth is an indication of the {\it scrambling} of the degrees of freedom probed by the operators. (Technically, the MSS bound concerns these quantum Lyapunov exponents, not the classical ones.) The most broadly accepted indicator of quantum chaos is the structure of a system's energy spectrum, where chaos manifests itself in the statistical distribution of energy eigenvalues. If the distribution approaches that of a random matrix theory (RMT) \cite{Wigner:1967ran,mehta2004random} within narrow energy bands, the system is conjectured to be chaotic \cite{Bohigas:1983er}. 

A related, but slightly different test of the thermalizing properties of a quantum system is the {\it Eigenstate Thermalization Hypothesis} (ETH) \cite{Deutsch:1991qu,Srednicki:1994mfb,DAlessio:2015qtq}. The ETH is concerned with the statistical properties of the matrix elements of an operator ${\cal A}$ in the energy representation:
\begin{equation}
\langle E_\alpha |{\cal A}| E_\beta \rangle 
= \langle {\cal A}(E) \rangle \delta_{\alpha\beta} +
e^{-S(E)/2} f_{\cal A}(E,\omega) R_{\alpha\beta} .
\label{eq:ETH}
\end{equation}
Here $E_\alpha$ are energy eigenstates and $E=(E_\alpha+E_\beta)/2$, $\omega=E_\alpha-E_\beta$. $S(E)$ is the microcanonical (mc) entropy and $\langle {\cal A}(E) \rangle$ the mc average of the observable ${\cal A}$, $f_{\cal A}(E,\omega)$ is called the spectral function for the operator ${\cal A}$. If the relation (\ref{eq:ETH}) holds, several statements can be proven:
\begin{itemize}
\item The long-time average ${\bar{\cal A}}$ approaches the thermal average $\langle{\cal A}\rangle_T$, implying that the system is ergodic.
\item The fluctuations of $\langle{\cal A}\rangle(t)$ around ${\bar{\cal A}}$ are exponentially small.
\item The quantum fluctuations of ${\cal A}$ represent the thermal fluctuations of the observable.
\item The real-time correlator of the observable ${\cal A}$ in an energy eigenstate satisfies
\begin{equation}
\langle E|{\cal A}(t){\cal A}(0)|E\rangle - \langle E|{\cal A}(t)|E\rangle\langle E|{\cal A}(0)E|\rangle \approx \int d\omega\,e^{-i\omega t}\, e^{\beta\omega/2} |f_{\cal A}(E,\omega)|^2 .
\end{equation}
\end{itemize}
In other words, when monitored through the observable ${\cal A}$, the quantum system in an energy eigenstate resembles a thermal system.

In order to show the validity of the ETH for the nonabelian gauge theory, the following steps need to be taken:
\begin{enumerate}
\item Discretize the continuum theory on a spatial lattice, choose boundary conditions, consider gauge invariant, multiplicatively renormalizable operators that are local or sufficiently smeared;
\item Show that the diagonal part is exponentially close to the microcanonical average;
\item Show that the off-diagonal part is a (Gaussian) random matrix;
\item Show that the spectral function decays exponentially for large $\omega$ and exhibits a (diffusive) plateau at small $\omega$;
\item Demonstrate renormalization group scaling for several lattice spacings $a$ while letting $g(a)\to 0$ to establish the continuum limit;
\item Demonstrate ETH for increasing system size at fixed $g(a)$ to establish the infinite volume limit. 
\end{enumerate}

To date, the steps described in the first four items have been successfully taken. The demonstration of the last two items will require greater computing resources and more powerful algorithms, or even quantum computers. We will now briefly discuss what is known for SU(2) gauge theory without fermions with respect to items $1-4$.

The gauge field Hamiltonian is discretized on a spatial lattice with the Kogut-Susskind (KS) Hamiltonian
\begin{equation}
H_{\rm KS} = \frac{g^2}{2} \sum_{\ell,a} \left(E_\ell^a\right)^2 + \frac{2}{g^2a^2} \sum_P (2-{\rm Tr}\,\square_P) ,
\label{eq:HKS}
\end{equation}
where the first sum is over lattice links $\ell$ and the second sum is over plaquettes $P$ and $g$ is the gauge coupling. $\square_P$ denotes the product of the gauge link variables $U(\ell,i)$ around the elementary plaquette $P$. Using the electric field basis \cite{Byrnes:2005qx}, the matrix elements of the electric term in the Hamiltonian are trivial, and those of the magnetic term (the plaquette term) can be expressed as products of Wigner 6-$j$ symbols \cite{Klco:2019evd}. For computational reasons, the electric basis must be truncated at a certain representation $j_{\rm max}$. Boundary conditions can be periodic or fixed by freezing all external links. So far, only lattices in two dimensions have been studied in detail; linear plaquette chains and area filling hexagonal (honeycomb) lattices \cite{Ebner:2023ixq}. In two dimensions the combination $g^2a$ is dimensionless and the gauge theory is super-renormalizable, which means that all quantities only depend on $g^2a$ and the continuum limit is also the weak-coupling limit.

\begin{figure}[ht]
\centering\includegraphics[width=0.45\linewidth]{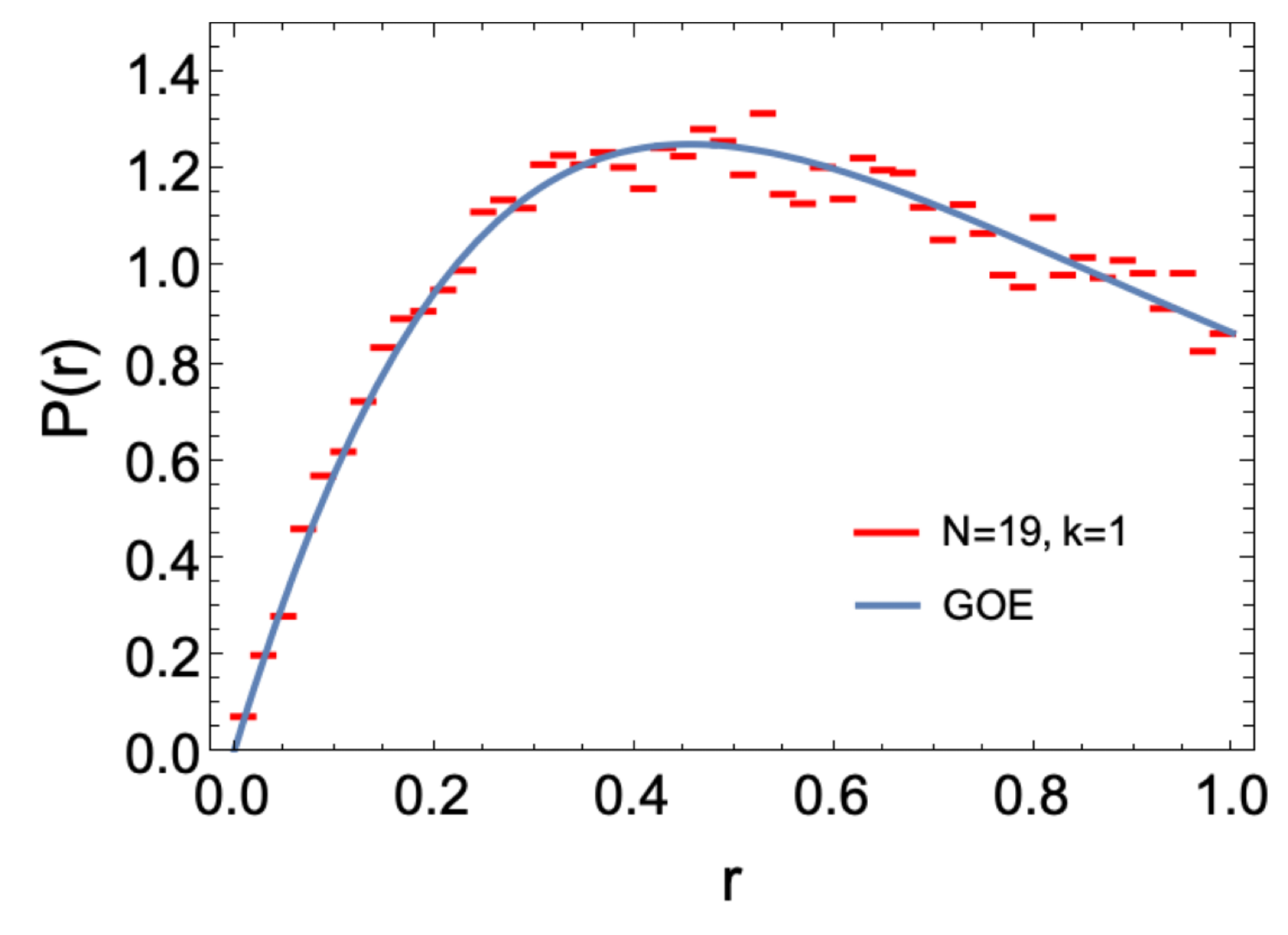}
\includegraphics[width=0.45\linewidth]{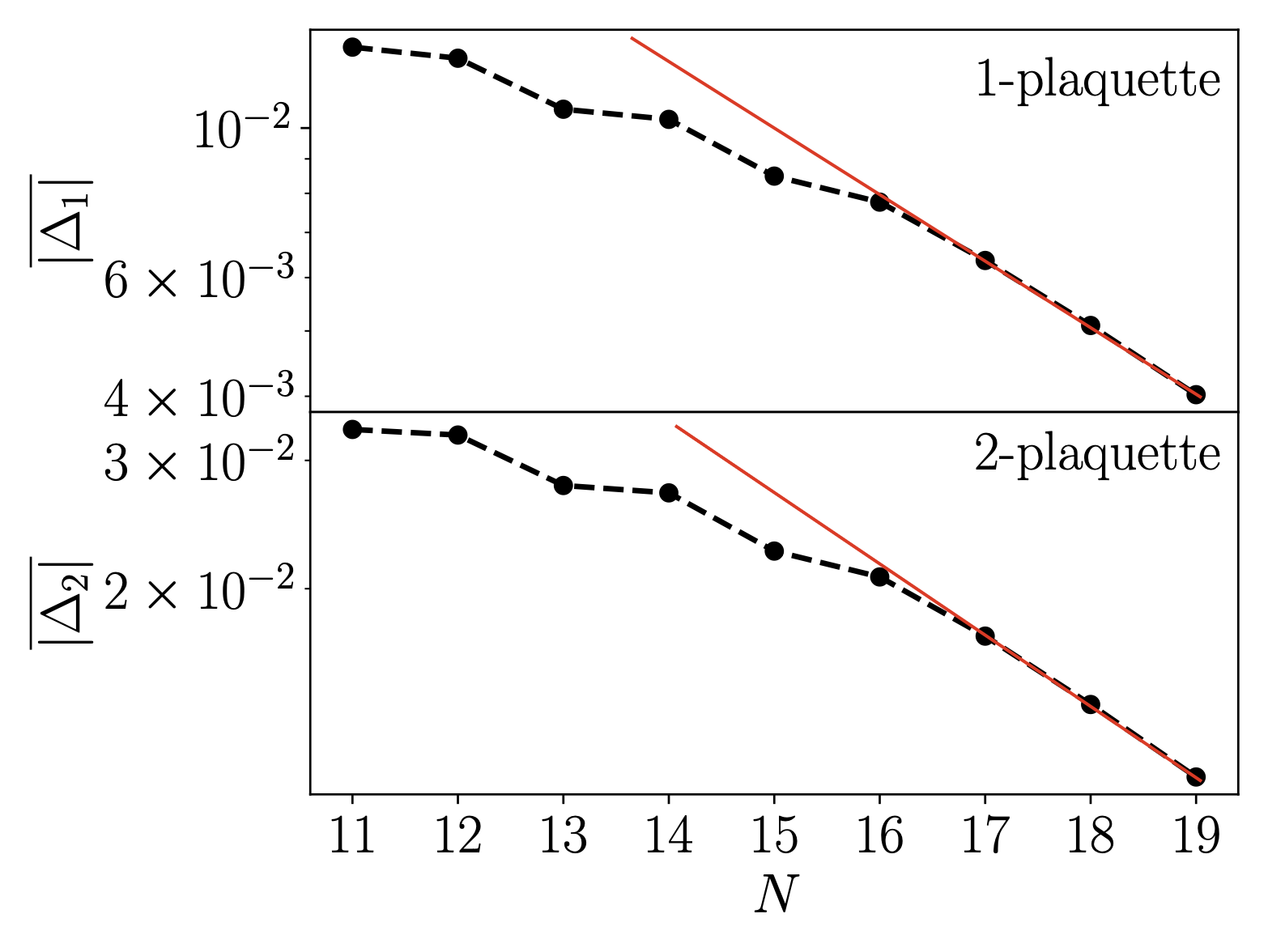}
\caption{Left panel: Restricted gap ratio distribution for energy levels for a $N=19$ plaquette chain with $j_{\max}=\frac{1}{2}$ and $g^2a=1.2$ in the region of highest level density (red dots) in comparison with the GOE prediction (solid blue line) \cite{Ebner:2023ixq}. 
Right panel: Average deviation $|\overline{\Delta_\nu}|$ of the diagonal matrix element of the one- and two-plaquette operators from their mc values for linear plaquette chains with $N=11-19$. The exponential fall-off is visible for $N\geq 16$ \cite{Yao:2023pht}.}
\label{fig:N19-plots}
\end{figure}

Computational resource constraints have so far limited the investigations to Hilbert spaces with dimension $10^5 - 10^6$. The number of basis states depends on the number of plaquettes $N$ and the cutoff $j_{\rm max}$. For linear plaquette chains, is has been possible to go to $N=19$ for $j_{\rm max} = \frac{1}{2}$, to $N=9$ for $j_{\rm max} = 1$, and to $N=3$ for $j_{\rm max} = 4$. For hexagonal lattices the largest size studied so far is $\vec{N} = (5,4)$ for $j_{\rm max} = \frac{1}{2}$. In all these instances the level spectrum within narrow energy windows has been found to closely resemble the Wigner-Dyson distribution for the Gaussian Orthogonal Ensemble (GOE) for moderate and weak coupling ($g^2a \lesssim 1$). Quantitative evidence comes from the nearest-neighbor level spacings, in particular, from the statistical distribution of the so-called restricted gap ratio that nicely agreed with the GOE prediction (see left panel of Fig.~\ref{fig:N19-plots}). 

\begin{figure}[ht]
\centering\includegraphics[width=0.45\linewidth]{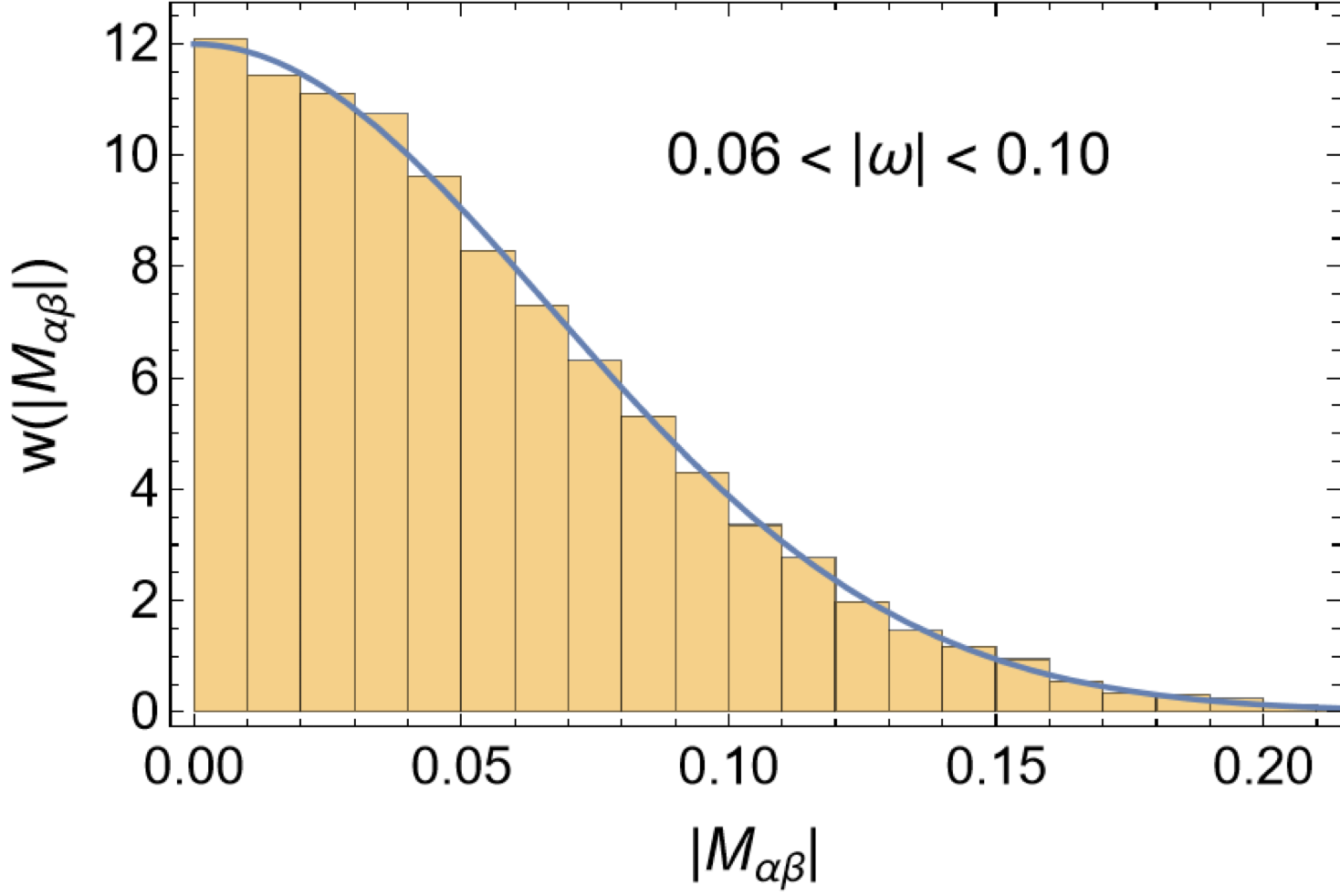}
\includegraphics[width=0.45\linewidth]{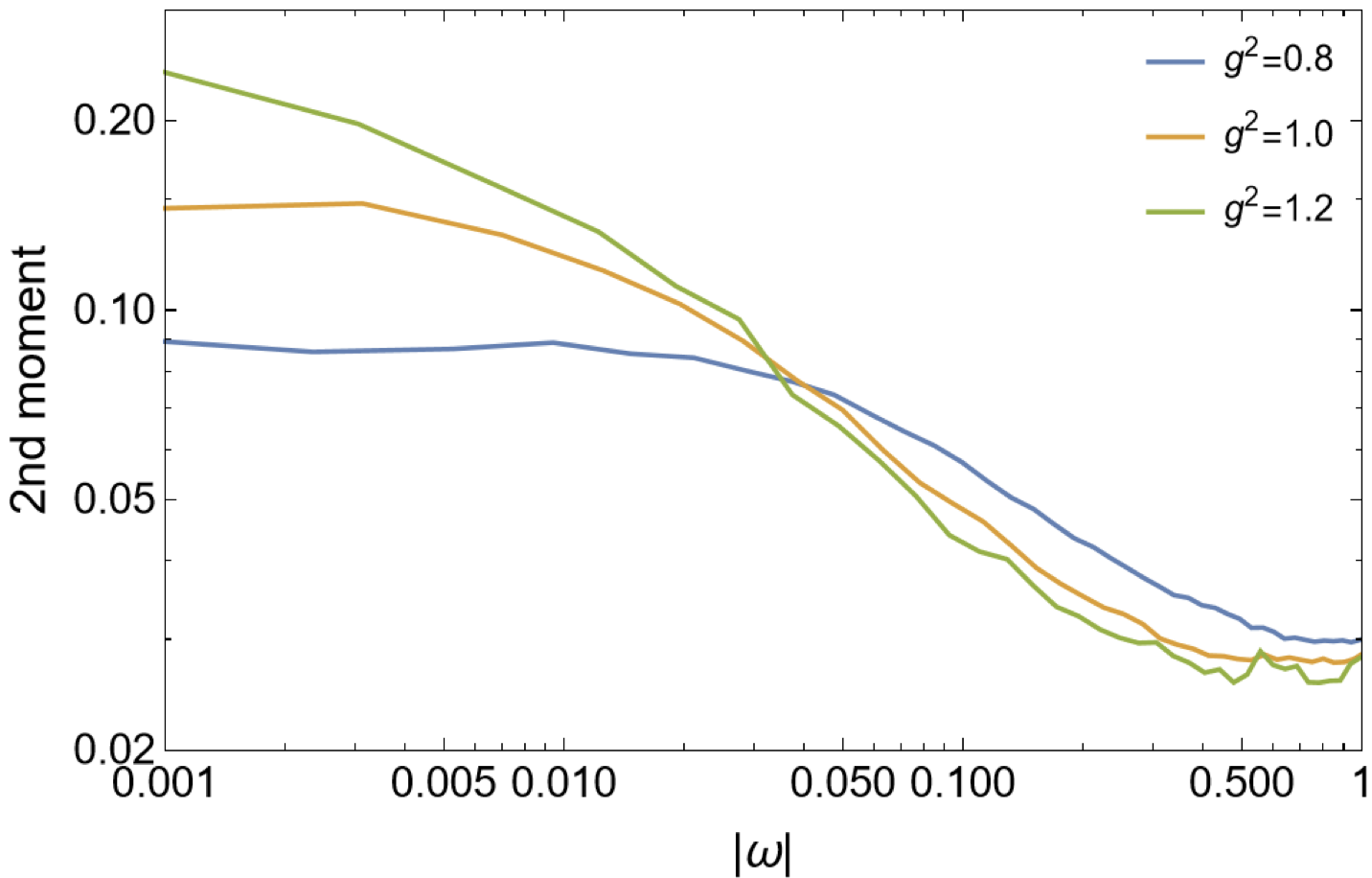}
\caption{Left panel: Statistical distribution of off-diagonal matrix elements for the $N=3$ plaquette chain for $j_{\rm max}=\frac{7}{2}$ in the energy difference window $0.06\leq |\omega|\leq 0.1$, together with a Gaussian fit \cite{Ebner:2023ixq}. 
Right panel: Spectral function for the same system for three different coupling constants \cite{Ebner:2023ixq}. As the coupling constant is reduced, the gauge theory enters more deeply into the chaotic regime at $g^2a \lesssim 1$ and the plateau at small $\omega$ becomes more prominent.}
\label{fig:N3-plots}
\end{figure}

With increasing lattice size $N$ the deviation of individual diagonal matrix elements of the one- and two-plaquette operators for the linear plaquette chains with $j_{\rm max}=\frac{1}{2}$ are found to exponentially converge with $N$ to the mc average (see right panel of Fig.~\ref{fig:N19-plots}). The off-diagonal matrix elements of the plaquette operator (magnetic energy) are found to follow Gaussian distributions \cite{Ebner:2023ixq}. The distribution in the energy difference window $0.06\leq |\omega|\leq 0.1$ for the $N=3$ chain with $j_{\rm max}=\frac{7}{2}$ is shown in the left panel of Fig.~\ref{fig:N3-plots}. The spectral function $f(E,\omega)$ can be deduced from the width of these Gaussian distributions; it is shown in the right panel of Fig.~\ref{fig:N3-plots} for the $N=3$ chain for three different coupling constants. It is clear that the plateau at small $|\omega|$ becomes more prominent as the coupling is reduced.

\section{Summary and Outlook}
\label{sec:summary}

In summary, recent numerical studies based on exact diagonalization of the KS Hamiltonian an small two-dimensional lattices have shown:
\begin{itemize}
\item Nearest neighbor level spacings obey GOE statistics.
\item Diagonal matrix elements of local operators cluster closely around the microcanonical average.
\item The off-diagonal matrix elements of local operators exhibit Gaussian random matrix behavior.
\item The spectral function exhibits a diffusive plateau at small $|\omega|$. 
\end{itemize}
Future investigations could pursue several targets including:
\begin{itemize}
\item (2+1)-dimensional hexagonal lattices \cite{Muller:2023nnk} with higher cutoff $j_{\rm max}$;
\item (3+1)-dimensional SU(2) gauge theory on a triamond lattice \cite{Kavaki:2024ijd};
\item SU(3) lattice gauge theory;
\item Gauge theories with fermions.
\end{itemize}
Finally, implementation of the KS Hamiltonian on a quantum computer \cite{Byrnes:2005qx,Klco:2019evd} could overcome the exponentially growing resource requirements on digital computers when and if computing platforms with a sufficient number of qubits and low noise level become available.

{\bf Acknowledgments:}
The authors gratefully acknowledge the scientific support and HPC resources provided by the Erlangen National High Performance Computing Center (NHR@FAU) of the Friedrich-Alexander-Universität Erlangen-Nürnberg (FAU) under the NHR project b172da-2. NHR funding is provided by federal and Bavarian state authorities. NHR@FAU hardware is partially funded by the German Research Foundation (DFG 440719683). BM acknowledges support from the U.S. Department of Energy, Office of Science, Nuclear Physics (awards DE-FG02-05ER41367). XY was supported by the U.S. Department of Energy, Office of Science, Office of Nuclear Physics, InQubator for Quantum Simulation (IQuS) (https://iqus.uw.edu) under Award Number DOE (NP) Award DE-SC0020970 via the program on Quantum Horizons: QIS Research and Innovation for Nuclear Science.

\bibliography{sn-bibliography}

\end{document}